\newcommand{\car}[4]{
\setlength{\unitlength}{1mm}
\makebox(20,20)[lb]{
\raisebox{-10mm}{
\put(5,5){\line(1,0){10}}
\put(5,5){\line(0,1){10}}
\put(5,15){\line(1,0){10}}
\put(15,5){\line(0,1){10}}

\put(2.5,2.5){\makebox(0,0){$#1$}}
\put(17.5,2.5){\makebox(0,0){$#2$}}
\put(17.5,17.5){\makebox(0,0){$#3$}}
\put(2.5,17.5){\makebox(0,0){$#4$}}
		}
	}
}
\newcommand{\carquatre}[4]{
\setlength{\unitlength}{1mm}
\makebox(20,20)[lb]{
\raisebox{-10mm}{
\put(5,5){\line(1,0){10}}
\put(5,5){\line(0,1){10}}
\put(5,15){\line(1,0){10}}
\put(15,5){\line(0,1){10}}
\put(15,15){\oval(10,10)[lb]}

\put(2.5,2.5){\makebox(0,0){$#1$}}
\put(17.5,2.5){\makebox(0,0){$#2$}}
\put(17.5,17.5){\makebox(0,0){$#3$}}
\put(2.5,17.5){\makebox(0,0){$#4$}}
          }
     }
}
\newcommand{\carun}[4]{
\setlength{\unitlength}{1mm}
\makebox(20,20)[lb]{
\raisebox{-10mm}{
\put(5,5){\line(1,0){10}}
\put(5,5){\line(0,1){10}}
\put(5,15){\line(1,0){10}}
\put(15,5){\line(0,1){10}}
\put(5,15){\oval(10,10)[rb]}

\put(2.5,2.5){\makebox(0,0){$#1$}}
\put(17.5,2.5){\makebox(0,0){$#2$}}
\put(17.5,17.5){\makebox(0,0){$#3$}}
\put(2.5,17.5){\makebox(0,0){$#4$}}
          }
     }
}
\newcommand{\cartrois}[4]{
\setlength{\unitlength}{1mm}
\makebox(20,20)[lb]{
\raisebox{-10mm}{
\put(5,5){\line(1,0){10}}
\put(5,5){\line(0,1){10}}
\put(5,15){\line(1,0){10}}
\put(15,5){\line(0,1){10}}
\put(15,5){\oval(10,10)[lt]}

\put(2.5,2.5){\makebox(0,0){$#1$}}
\put(17.5,2.5){\makebox(0,0){$#2$}}
\put(17.5,17.5){\makebox(0,0){$#3$}}
\put(2.5,17.5){\makebox(0,0){$#4$}}
          }
     }
}
\newcommand{\cardeux}[4]{
\setlength{\unitlength}{1mm}
\makebox(20,20)[lb]{
\raisebox{-10mm}{
\put(5,5){\line(1,0){10}}
\put(5,5){\line(0,1){10}}
\put(5,15){\line(1,0){10}}
\put(15,5){\line(0,1){10}}
\put(5,5){\oval(10,10)[rt]}

\put(2.5,2.5){\makebox(0,0){$#1$}}
\put(17.5,2.5){\makebox(0,0){$#2$}}
\put(17.5,17.5){\makebox(0,0){$#3$}}
\put(2.5,17.5){\makebox(0,0){$#4$}}
          }
     }
}

\newcommand{\carcdeux}[4]{
\setlength{\unitlength}{1mm}
\makebox(20,20)[lb]{
\raisebox{-10mm}{
\put(5,5){\line(1,0){10}}
\put(5,5){\line(0,1){10}}
\put(5,15){\line(1,0){10}}
\put(15,5){\line(0,1){10}}
\put(15,15){\oval(10,10)[lb]}
\put(5,5){\oval(10,10)[tr]}

\put(2.5,2.5){\makebox(0,0){$#1$}}
\put(17.5,2.5){\makebox(0,0){$#2$}}
\put(17.5,17.5){\makebox(0,0){$#3$}}
\put(2.5,17.5){\makebox(0,0){$#4$}}
          }
     }
}
\newcommand{\carcun}[4]{
\setlength{\unitlength}{1mm}
\makebox(20,20)[lb]{
\raisebox{-10mm}{
\put(5,5){\line(1,0){10}}
\put(5,5){\line(0,1){10}}
\put(5,15){\line(1,0){10}}
\put(15,5){\line(0,1){10}}
\put(5,15){\oval(10,10)[rb]}
\put(15,5){\oval(10,10)[tl]}

\put(2.5,2.5){\makebox(0,0){$#1$}}
\put(17.5,2.5){\makebox(0,0){$#2$}}
\put(17.5,17.5){\makebox(0,0){$#3$}}
\put(2.5,17.5){\makebox(0,0){$#4$}}
          }
     }
}

\documentstyle[A4,12pt,epsf]{article}
\begin{document}

\title{ON THE CONSTRUCTION OF INTEGRABLE DILUTE A-D-E MODELS.}
\author{{\sc Ph. ROCHE}\thanks{On leave from
Centre de Physique Th\'eorique, Ecole Polytechnique
91128 PALAISEAU Cedex, FRANCE.
Email: PTHPRO@frpoly11.bitnet}\\
Laboratoire de Physique Th\'eorique ENSLAPP\\
Ecole Normale Sup\'erieure de Lyon\\
46 All\'ee d'Italie, 69007 LYON, FRANCE.\date{Mars 1992}}

\maketitle

\abstract{
We give an integrable extension of the lattice models recently considered
by I.~Kostov in his study of strings in discrete space. These models are
IRF models with spins variables living in any connected graph, the vertex
model underlying these models being the Izergin-Korepin model.  When the
graph is taken to be the  simply laced Dynkin Diagrams, it is conjectured
that these models  possess  critical regimes which are the dilute phase of
SOS models of ADE type. }

\section{Introduction}

During these last few years there has been an incredible amount of work
studying the connections
between  integrable lattice models and conformal field theory. One
particularly fruitful but difficult problem consists of finding in each
universality class of critical phenomena an integrable spin model.
In the case where  the central charge satisfies $c<1$ this  problem is almost
completely understood. The aim of this letter is to provide the still
missing unitary integrable lattice model having $c<1$.

There exists a  classification of unitary conformal field theory of central
charge  $c<1$ [CIZ] based on a complete list of modular invariant partition
functions. The central charge of these CFT are equal to $c=1-6
{(p-p')^2\over pp'}$ with $\vert p-p'\vert=1.$ The modular invariants are
classified by a pair $(A_{p-1},G_{p'-1})$ with $G$ being a simple Lie
algebra of $A, D, E$ type, $(p,p')$ indexing the largest exponents of the
corresponding Lie algebras.

V.Pasquier has shown [Pa1] that it is possible to generalize  the
construction of RSOS models of Andrews-Baxter-Forrester [ABF] to any simply
laced Dynkin Diagram.
The definition of these models is as follows. Let $(C_{ij})$ denote a
symmetric matrix with coefficients equal to 0 or 1, this matrix define a
graph: points of the graph are the  entries of the matrix and two points
$i,j$ are connected by a line of the graph if the matrix element $C_{ij}$
is different of zero. We will suppose that the corresponding graph $G$ is
connected. In that case the Perron Frobenius theorem asserts that there
exists a greatest positive eigenvalue $\beta$ of $C$ which is non
degenerate and that the corresponding eigenstate can be chosen such that
all is components $(S_i)$ are positive.  The spin variables of this model
are points of the graph $G$, the Boltzmann weights associated to a
configuration of spins surrounding a plaquette is defined by:
\begin{equation}
W(a,b,c,d)=\delta_{b,d}+ {X\over \beta}{(S_b S_d)^{1/2}\over S_a}
\delta_{a,c}
\end{equation}

with the restriction that $(a,b)$  is a couple of connected points on the
graph, as well as $(b,c)$, $(c,d)$ and $(d,a)$.
We can represent the Boltzmann weight by the following picture
\begin{equation}
W(a,b,c,d)=\carcun{b}{c}{d}{a}+\carcdeux{b}{c}{d}{a}
\end{equation}
where the lines inside the square separate connected spins on the graph
$G$.

These weights satisfy Yang-Baxter equation, i.e the equation:
\begin{eqnarray}
\lefteqn{\sum_{g} W''(a_1, a_2, a_3, g) W'(g, a_3, a_4, a_5) W(a_1, g, a_5,
a_6)
=} \nonumber \\
  &  & \sum_{g} W(a_2, a_3, a_4, g) W'(a_1, a_2, g, a_6) W''(a_6, g, a_4,
a_5)
\end{eqnarray}
provided that the relation  $X'(1-{1\over \beta}XX'')=X+X''+XX''$ is
satisfied.

\medskip

When $X=\beta$ this model is  isotropic and is in a critical regime  when
$\beta\leq 2.$ This last constraint implies that the graph is a Dynkin
Diagram of ADE type or an extended Dynkin diagram $\hat A\hat D\hat E.$
It can be shown [Pa2] that the partition function  on a torus  of the
continuum limit of these models are the modular invariants of
$(A_{p-1},G_{p'-1})$ type with $p<p'$. There are strong evidence that the
statistical models associated to the other branch $p>p'$ can be described
by a dilute version of these ADE SOS models [Ni].
An important step has been pushed forward by I.Kostov in  [Ko1]. In this
work he defines a class of statistical models where spins are points of a
graph $G$ and the Boltzmann weights are taken to be :

$$
W(a,b,c,d)=\car{b}{c}{d}{a}+\carcun{b}{c}{d}{a}+\carcdeux{b}{c}{d}{a}
$$

\begin{equation}
\hfill\carun{b}{c}{d}{a}+\cardeux{b}{c}{d}{a}+\cartrois{b}{c}{d}{a}+
\carquatre{b}{c}{d}{a}
\end{equation}
with
\begin{equation}
\carcun{b}{c}{d}{a}={1\over T^2}\delta_{b,d}C_{ab}C_{bc},
\carcdeux{b}{c}{d}{a}={1\over T^2}\delta_{a,c}C_{ab}C_{ad}({{S_b S_d}\over
{S_a S_c}})^{1/2}
\end{equation}

\begin{equation}
\carun{b}{c}{d}{a}={1\over T}\delta_{b,c}\delta_{c,d}C_{ab},
\cartrois{b}{c}{d}{a}={1\over T}\delta_{a,b}\delta_{a,d}C_{bc}
\end{equation}

\begin{equation}
\cardeux{b}{c}{d}{a}={1\over T}\delta_{a,d}\delta_{d,c}C_{ab}{({S_b\over
S_a})}^{1/2},
\carquatre{b}{c}{d}{a}={1\over T}\delta_{a,b}\delta_{b,c}C_{cd}{({S_d\over
S_a})}^{1/2}
\end{equation}

\begin{equation}
\car{b}{c}{d}{a}=\delta_{a,b}\delta_{b,c}\delta_{c,d}
\end{equation}

where T is the temperature of the statistical system.
When T goes to zero the Boltzmann weights are those of the type described by
(1), when T
goes to infinity the configuration of spins freeze to a constant and there
are no long range correlations. He expects that there exists a critical
temperature $T_c$ at which there is a phase transition between these two
phases. At this point there is a new critical regime, which is called the
dilute phase [Ni].

 He argues that the partition function of these models at
$T_c$ is indeed the modular invariants  $(A_{p-1},G_{p'-1})$ with $p>p'$.
Unfortunately the Boltzmann weights are not written in a way that could
satisfy Yang-Baxter equation, one problem being that there is no spectral
parameter.
In this letter we describe a family of integrable models such that the
Boltzmann weights at the isotropic point  is of the type (1).

\section{Description of the solution}

We have searched a solution of Yang Baxter equation of the type:

$$
W(a,b,c,d)=A \delta_{a,b}\delta_{b,c}\delta_{c,d}+
C_1\delta_{b,d}C_{ab}C_{bc}+
C_2\delta_{a,c}C_{ab}C_{ad}({{S_b S_d}\over {S_a S_c}})^{1/2}+
$$
$$
B_1\delta_{b,c}\delta_{c,d}C_{ab}+
B_3
\delta_{a,b}\delta_{a,d}C_{bc}+B_2\delta_{a,d}\delta_{d,c}C_{ab}{({S_b\over
S_a})}^{1/2}+B_4 \delta_{a,b}\delta_{b,c}C_{cd}{({S_d\over S_a})}^{1/2}+
$$
\begin{equation}
D_1 \delta_{a,b}\delta_{cd} C_{bc}+
D_2 \delta_{a,d}\delta_{b,c}C_{ab}
\end{equation}

The coefficients $A$, $B_1$, $B_2$, $B_3$, $B_4$, $C_1$, $C_2$,
$D_1$, $D_2$ being  unknown variables.  We have taken a slighter
generalisation of the weights (5-8) by allowing new types of Boltzmann
weights ($D_1$ and $D_2$) otherwise there is no interesting solution.
Notice that these  Boltzmann weights do not depend on the spin variables.
After having written the whole family of Yang-Baxter equations, one
discovers that all factors depending on $S_i$ diseappear and combine nicely
leaving sometimes a coefficient $\beta$.
The system of algebraic  equations one obtains consits in  49  equations.
The easiest to handle are the four
\begin{eqnarray}
B_1 B_3' B_1'' & = & B_3 B_1' B_3''\\
B_2 B_4' B_1'' & = & B_4 B_2' B_3''\\
B_3 B_4' B_2'' & = & B_1 B_2' B_4''\\
B_4 B_3' B_2'' & = & B_2 B_1' B_4''
\end{eqnarray}
from which one deduces that $B_1=B_3$ and $B_2=\Gamma B_4 $  with $\Gamma$
constant.

\medskip

The whole set of algebraic equations reduces then to :
\begin{eqnarray}
-B_1 C_1' B_1''+ C_1 B_1' C_1''-D_2 B_1' D_1'' & = & 0 \\
-B_2 D_2' B_1''+D_2 B_2' C_1''-C_2 B_2'D_1'' & = & 0   \\
-D_1 C_1' B_1''+C_1 D_1'B_1''-B_1 B_1' D_1'' & = & 0  \\
-D_2 B_1' B_1''+B_1 D_2' C_1''-B_1 C_1' D_2'' & = & 0 \\
-C_2 B_2' B_1''+B_2 C_2' C_1''-B_2 D_2' D_2'' & = & 0  \\
-D_1 C_2' B_1''+C_2 D_1' B_1''+B_4 B_2' D_2'' & = & 0  \\
-D_1 B_2' B_4'' -B_1 D_2' C_2''+B_1 C_2' D_2'' & = & 0 \\
B_1 D_1' B_2''+D_2 B_2' C_2''- C_1 B_2' D_1'' & = & 0 \\
-C_1 C_2' B_2''+D_1 D_1' B_2''+B_1 B_2' C_2'' & = & 0
\end{eqnarray}
\begin{eqnarray}
D_2 A' B_1''-A D_2' B_1''-B_4 B_2' D_1'' +B_1 B_1'D_2'' & = & 0 \\
-D_1 B_2' A''+B_2 D_1' B_1''-B_1 D_2' B_2''+A B_2' D_2'' & = & 0 \\
B_1 D_1' A''-D_1 B_1' B_1''+D_2 B_2' B_4''-B_1 A' D_1'' & = & 0
\end{eqnarray}
\begin{equation}
-B_2 B_1' B_4'' +C_1 C_2' C_1''
-\beta C_2 C_1' C_2''-C_2 C_2' C_2''-C_2 C_1'C_1''-C_1 C_1' C_2''=0
\end{equation}
\begin{eqnarray}
-B_2 B_1' A''-C_1 C_1' B_2''-C_2 C_2' B_2''+B_1 B_2' C_1''-\beta C_2 C_1'
B_2'' & = & 0  \\
C_1 B_4' B_1''- A B_1' B_4''-\beta B_4 C_1' C_2''-B_4 C_2' C_2''-B_4 C_1'
C_1'' & = & 0  \\
-A B_1' A''+B_1 A' B_1''-\beta B_4 C_1' B_2''-B_4 C_2' B_2''+D_1 B_1' D_2''
& = & 0 \\
-B_1 C_2' B_1''+B_4 A' B_2''+\beta C_2 B_1'C_2''+C_2 B_1' C_1''+C_1 B_1'
C_2'' & = & 0 \\
-B_1 B_2' A''+ A A' B_2''- D_2 D_2' B_2'' +\beta B_2 B_1'C_2''+B_2 B_1'
C_1'' & = & 0  \\
-B_2 A' A''- C_1 B_1' B_2''-\beta C_2 B_1' B_2''+ A B_2'B_1''+B_2 D_1'D_1''
& = & 0
\end{eqnarray}
One solves this system through the following steps.

We first find the invariants, i.e rational functions of the unknowns which
are equal for the unprime variables and for the prime variables. These
functions will define an algebraic manifold whose points are the spectral
parameter.

By eliminating the variables with double primes in the system (14, 15, 16)
and (17, 18, 19), we obtain that:
\begin{equation}
\theta_2={{C_1 C_2 -D_2^2}\over B_1 B_2}
\end{equation}
\begin{equation}
\Omega_2={B_1 D_2\over C_1 B_2}+
\theta_2 {D_1\over C_1}\ {\rm and}\  \tilde\Omega_2={B_2 D_2\over C_2 B_1}+
\Gamma \theta_2{D_1\over C_2}
\end{equation} are invariants.

By eliminating the unprime variables in the system (16, 20, 22) and (14,
17, 21), we obtain that:
\begin{equation}
\theta_1={{C_1 C_2 -D_1^2}\over B_1 B_2}
\end{equation}
\begin{equation}
\Omega_1={B_1 D_1\over C_1 B_2}+
\theta_1 {D_2\over C_1}\ {\rm and}\  \tilde\Omega_1={B_2 D_1\over C_2 B_1}+
\theta_1\Gamma  {D_2\over C_2}
\end{equation}
 are invariants.

By eliminating the prime variables in the system (15, 18, 20) and (19, 21,
22), we obtain that:
\begin{equation}
\Omega_2 \Gamma={B_2 D_1\over C_2 B_1}+
\theta_2 \Gamma {D_2\over C_2}\ {\rm and}\  \Omega_1 \Gamma ={B_2 D_2\over
C_2 B_1}+
 \Gamma \theta_1 {D_1\over C_2}
\end{equation}
  are invariants.

The only interesting solution corresponds to $D_1=D_2=D$ from which we
deduce that:

\begin{equation}
C_1 C_2-D^2= \theta_2 B_1 B_2
\end{equation}

\begin{equation}
{D\over C_1}(\theta_2 +{B_1\over B_2})={D\over C_2}(\theta_2 +\Gamma^{-1}
{B_2\over B_1})=\Omega_2
\end{equation}
We can now eliminate the double prime variables in (20,21,22) and one must
have that
\begin{equation}
{{C_2^2+D^2-\Gamma^{-1} B_2^2}\over 2C_2 D}={{C_1^2+D^2-B_1^2}\over 2 D
C_1}=\Delta.
\end{equation}

As usual one can put $\Delta={1\over 2}(q^2+q^{-2})$ and
one is led to parametrize
\begin{eqnarray}
D&=&\rho_1 (x_1-x_1^{-1})=\rho_2 (x_2-x_2^{-1})\\
 C_1&=&\rho_1 (q^2 x_1- q^{-2}x_1^{-1}),\\
 C_2&=&\rho_2 (q^2 x_2- q^{-2}x_2^{-1})\\
 B_1&=&\rho_1 (q^2 - q^{-2}),
B_2=\Gamma^{1/2}\rho_2 (q^2 - q^{-2})
\end{eqnarray}
The equations (14...22) are then equivalent to the fact that  the product
$x_1 x_2$ is fixed and equal to a constant $\lambda,$ function of the
invariant $\Delta, \Gamma , \Omega_1. $  Defining $x=x_1$ the system
(14...22) is now equivalent to the simple relation   $x'=x x''.$
Finally equation (26) implies that $\lambda= q^{-3},$ and
$\beta=-(q^4+q^{-4}).$
$A$ is finally obtained by solving (27).

The final result is the parametrisation of weights up to a constant :

\begin{eqnarray}
A(x)&=&q^3 x^2 +q^{-3}x^{-2}+ (q+q^{-1})(1-q^4-q^{-4})\\
B_1(x)&=&B_3(x)=(q^2-q^{-2})(q^{-3} x^{-1}-q^3 x)\\
\Gamma^{1/2} B_2(x)&=&\Gamma^{-1/2} B_4(x)=(x-x^{-1})(q^2-q^{-2})\\
C_1(x)&=&(q^2 x -q^ {-2} x^{-1})(q^{-3}x^{-1}- q^3 x)\\
C_2(x)&=&( x - x^{-1})(q^{-1}x^{-1}- q x)\\
D_1(x)&=&D_2(x)=(x-x^{-1})(q^{-3}x^{-1}-q^3 x).
\end{eqnarray}

It is then quite easy to show that Yang-Baxter  equations are satisfied
using this parametrisation.

Remark: we have discarded the solution corresponding to $D_1=D_2=0$
otherwise we would have obtained
 the ordinary SOS model (1) with an adjacency matrix equal to $G+1$ which
clearly has always $\beta>2$
and which do not posess a critical regime.

We have then obtained an integrable IRF model,
 which we will call diluted, associated to any graph   $G.$ This is very
reminiscent of the
construction of the  RSOS models (1) sketched in the introduction.
 It is well known that the ordinary $A_n$ models can be obtained by a
vertex -IRF transformation
from the 6 vertex model.
 One can ask the same question in the case of these diluted model when
associated to $A_n$ Dynkin
diagrams. The vertex model one obtains  is the Izergin-Korepin model [IK].
As shown by M.Jimbo [Ji]
the R matrix of the Izergin- Korepin model is the R matrix in the
fundamental representation (of
dimension 3) of the deformation of the Kac-Moody algebra $U_k(A_2^{(2)}).$
If we set $k=-q^2$ in the
parametrisation of M.Jimbo we obtain up to a normalization constant the
vertex model one can
straightforwardly obtain from the Boltzmann weights in the $A_n$ case we
have computed (one just has
to get rid of the factor $S_a$).

\section{Conclusion}
Although our study has just been an algebraic one,
what remains to be done is the study of the thermodynamical properties of these
models.
One has to  study  the critical regimes of these models and give a proof
that when the graph $G$ is
taken to be a Dynkin diagram of $A D E$ type then there exists a critical
regime where it is described
by the minimal conformal field theories in the branch $p>p.'$ We would
then have a one to one
correspondance between minimal unitary CFT and a family of lattice
integrable models.

The study of the perturbed minimals models by $\phi_{1,2}$ has shown that
the $S$-matrix of these
fields theories are the RSOS version of the R-matrix of Izergin-Korepin
[Sm].  It would be
interesting   to understand the  connection, if any, between these field
theories and the lattice
models we have built.

{}From a pure  mathematical point of view the ordinary ADE models provide
a natural representation of the Temperley-Lieb algebra on the space of
paths of any graph $G$. As
shown by V.Jones in [Jo], this algebra is a central tool in the study of
the inclusion of hyperfinite
factors. Because our construction of integrable models works as well for
any graph $G$ it is rather
natural to wonder if these models can have a natural interpretation in
terms of inclusion of factors.

\medskip
Note added:

 After the completion of this work we received a preprint of S.O.Warnaar,
B.Nienhuis and K.A.Seaton
[WNS] where they give the same formula (9) in a different parametrisation.
In order to get their
weights from (45,50) one has to first take the opposite of $D$ and then put
$q=i e^{i\lambda}$ and
$x=e^{i u}.$  They have obtained an off critical extension of $A_n$ models,
and they give
results on the study of the critical regimes of the statistical models
defined by (9).

\medskip
{\bf Acknowledgments}

We would like to thank I.Kostov  for numerous discussions,
 D.Bernard for pointing out the relevance  of  the  Izergin-Korepin
solution in our problem and
people of the ENSLAPP who generate a very stimulating atmosphere.

\section{References}
\item{[ABF]}{ Andrews.G.E, Baxter.R.J, Forrester.P.J, {\sl Eight vertex
SOS model and generalized Rogers-Ramanujan type identities },
Jour.Stat.Phys {\bf 35}, 193 (1984)}

\item{[CIZ]} { Capelli.A, Itzykson.C, Zuber.J.B:{\sl The ADE classification
of minimal and $A^{(1)}$ conformal invariant theories},   Commun.Math.Phys.
{\bf 113},1 (1987).}

\item{[IK]}{ Izergin, A.G, Korepin.V.E:{\sl The inverse scattering approach
to the quantum Shabat-Mikhailov model}, Commun.Math.Phys. {\bf 79}, 303
(1981).}
\item{[Jo]}{ Jones.V.F.R,{\sl Index for subfactors}, Invent.Math {\bf 72},
1 (1983).}

\item{[Ji]} { Jimbo.M: {\sl Quantum R matrix for the generalized Toda
system},  Commun.Ma\-th.Phys.{\bf 10},63 (1985).}
\item{[Ko1]}{ Kostov.I:{\sl Strings in discrete target space},  Saclay
Preprint SphT,91-142.}

\item{[Ni]}{ Nienhuis.B:{\sl Analytical calculation of two leading
exponents of the dilute Potts model}, J.Phys.A.Math.Gen {\bf 15},199
(1982).}
\item{[Pa1]}{ Pasquier.V: {\sl two dimensional critical system labelled by
Dynkin diagrams}, Nuc.Phys B {\bf 285} [FS 19], 162 (1986).}
\item{[Pa2]}{Pasquier.V:  {\sl  Lattice derivation of modular invariant
partition functions on the torus}, J.Phys.A. Math.Gen 20 L 1229 (1987)}
\item{[Sm]}{Smirnov.F.A: {\sl Exact S-Matrix for $\phi_{1,2}$-perturbed
minimal models of CFT},
Int.Jour.Mod.Phys. A, Vol 6, N 8, 1407, (1991)} \item{[WNS]}{Warnaar.S.O,
Nienhuis.B, Seaton.K.A: {\sl
A New Construction of Solvable RSOS Models Including an Ising Model in a
Field}, Instituut voor
Theoretische Fysica, Universiteit van Amsterdam Preprint (1992)}

\end{document}